\input harvmac

\def \A {{\cal A}}

\def\vp{{\bf p}}

\def \ep{\epsilon}

\def \om {\omega}

\def\be{\begin{equation}}
\def\ee{\end{equation}}

\def \k {\kappa} 
\def \F {{\cal F}}
\def \g {\gamma}
\def \del {\partial}
\def \bd {\bar \partial }

\def \ha{{\textstyle{1\over 2}}}

\def \a {\alpha}
\def \b {\beta}
\def \chi {\chi}\def\r {\rho}
\def \s {\sigma}
\def \p {\phi}
\def \m {\mu}
\def \n {\nu}
\def \vp {\varphi }
\def \l {\lambda}
\def \t {\theta}
\def \td {\tilde }
\def \d {\delta}

\def \sm {$\s$-model }

\def \P {\Phi}

\def \inv {^{-1}}
\def \ov {\over }

\def \sin{{\rm sin}}
\def \cos{{\rm cos}}

\def \lr { \lref}
\def\np {{  Nucl. Phys. }}
\def \pl {{  Phys. Lett. }}
\def \mpl {{ Mod. Phys. Lett. }}
\def \prl {{  Phys. Rev. Lett. }}
\def \pr  {{ Phys. Rev. }}

\def \cqg {{ Class. Quant. Grav. }}

\baselineskip8pt
\Title{
\vbox
{\baselineskip 6pt{\hbox{  }}{\hbox
{Imperial/TP/95-96/48}}{\hbox{hep-th/9605091}} {\hbox{
  }}} }
{\vbox{\centerline {Extremal black hole  entropy  }
\vskip2pt 
\centerline {from conformal string sigma model }
\vskip4pt
}}
\vskip -20 true pt



\centerline{   A.A. Tseytlin\footnote{$^{\star}$}{\baselineskip8pt
e-mail address: tseytlin@ic.ac.uk}\footnote{$^{\dagger}$}{\baselineskip8pt
On leave  from Lebedev  Physics
Institute, Moscow.} }

\smallskip\smallskip
\centerline {\it  Theoretical Physics Group, Blackett Laboratory,}
\smallskip

\centerline {\it  Imperial College,  London SW7 2BZ, U.K. }
\bigskip\bigskip
\centerline {\bf Abstract}
\medskip
\baselineskip10pt
\noindent
We present string-theory derivation of the semiclassical
entropy of extremal dyonic black holes in the approach based
on conformal sigma model (NS-NS embedding of the classical
solution). We demonstrate (resolving some puzzles existed
in previous related discussions) that the degeneracy
responsible for the entropy is due to string oscillations
in four transverse dimensions `intrinsic' to black hole:
four non-compact directions of the D=5 black hole and
three non-compact and one compact (responsible for embedding
of magnetic charges) dimensions in the D=4 black hole case.
Oscillations in  other compact internal dimensions give
subleading contributions to  statistical entropy in the limit
when all charges are large. The dominant term in the statistical
entropy is thus universal (i.e. is the same in type II
and heterotic string theory) and agrees with Bekenstein-Hawking
expression.

\medskip
\Date {May 1996}
\noblackbox
\baselineskip 14pt plus 2pt minus 2pt
\lr \duh {A. Dabholkar, G.W. Gibbons, J. Harvey and F. Ruiz Ruiz,  \np
B340 (1990) 33;
A. Dabholkar and  J. Harvey,  \prl
63 (1989) 478.
}
\lr\mon{J.P. Gauntlett, J.A. Harvey and J.T. Liu, \np B409 (1993) 363.}
\lr\chs{C.G. Callan, J.A. Harvey and A. Strominger, 
\np { B359 } (1991)  611; in {\it 
Proceedings of the 1991 Trieste Spring School on String Theory and
Quantum Gravity}, J.A. Harvey {\it et al.,}  eds. (World Scientific, 
Singapore
1992).}

\lr\MV {J.C. Breckenridge, R.C. Myers, A.W. Peet  and C. Vafa, HUTP-96-A005,  hep-th/9602065.}

\lr \peet {A.W.  Peet, \np  {B456} (1995) 732, 
 hep-th/9506200.}

\lr \US{M. Cveti\v c and  A.A.  Tseytlin, 
\pl { B366} (1996) 95, hep-th/9510097. 
}
\lr \USS{M. Cveti\v c and  A.A.  Tseytlin, 
\pr D53 (1996) 5619, hep-th/9512031. 
}
\lr\LW{ F. Larsen  and F. Wilczek, 
 hep-th/9511064;   hep-th/9604134.}
\lr\LWW{ F. Larsen  and F. Wilczek, 
    hep-th/9604134.}
\lr\horr{G.T. Horowitz, hep-th/9604051.}

\lr\TT{A.A. Tseytlin, \mpl A11 (1996) 689, hep-th/9601177.}
\lr \HT{ G.T. Horowitz and A.A. Tseytlin,  \pr { D51} (1995) 
2896, hep-th/9409021.}
\lr\khu{R. Khuri, \np B387 (1992) 315; \pl B294 (1992) 325.}
\lr\CY{M. Cveti\v c and D. Youm,
 UPR-0672-T, hep-th/9507090; UPR-0675-T, hep-th/9508058; 
  \pl { B359} (1995) 87, 
hep-th/9507160.}

\lr \CM{ C.G. Callan and  J.M.  Maldacena, 
  hep-th/9602043.} 
\lr\SV {A. Strominger and C. Vafa,   hep-th/9601029.}

\lr\MV {J.C. Breckenridge, R.C. Myers, A.W. Peet  and C. Vafa, HUTP-96-A005,  hep-th/9602065.}
\lr\vijay{V. Balasubramanian and F. Larsen, hep-th/9604189.}
\lr \US{M. Cveti\v c and  A.A.  Tseytlin, 
\pl { B366} (1996) 95, hep-th/9510097. 
}
\lr\LW{ F. Larsen  and F. Wilczek, 
  hep-th/9511064.    }

\lr\DasMat{S. Das and S. Mathur, hep-th/9601152.}

\lr\mon{J.P. Gauntlett, J.A. Harvey and J.T. Liu, \np B409 (1993) 363.}

\lr\mina{M.J. Duff, J.T. Liu and R. Minasian, 
\np B452 (1995) 261, hep-th/9506126.}
\lr\dvv{R. Dijkgraaf, E. Verlinde and H. Verlinde, hep-th/9603126;
hep-th/9604055.}
\lr\gibb{G.W. Gibbons and P.K. Townsend, \prl  71
(1993) 3754, hep-th/9307049.}
\lr\town{P.K. Townsend, hep-th/9512062.}
\lr\kap{D. Kaplan and J. Michelson, hep-th/9510053.}

\lr\ast{A. Strominger, hep-th/9512059.}
\lr \ttt{P.K. Townsend, hep-th/9512062.}
\lr \papd{G. Papadopoulos and P.K. Townsend, hep-th/9603087.}
\lr\jch {J. Polchinski, S. Chaudhuri and C.V. Johnson, 
hep-th/9602052.}

\lr \gig{G.W. Gibbons, M.J. Green and M.J. Perry, 
hep-th/9511080.}

\lr \US{M. Cveti\v c and  A.A.  Tseytlin, 
\pl {B366} (1996) 95, hep-th/9510097.  
}
\lr\mast{J.M. Maldacena and A. Strominger, hep-th/9603060.}
\lr \CY{M. Cveti\v c and D. Youm,
 \pr D53 (1996) 584, hep-th/9507090.  }
 \lr\kall{R. Kallosh, A. Linde, T. Ort\' in, A. Peet and A. van Proeyen, \pr { D}46 (1992) 5278.} 
\lr \grop{R. Sorkin, Phys. Rev. Lett. { 51 } (1983) 87;
D. Gross and M. Perry, Nucl. Phys. { B226} (1983) 29. 
}
\lr \myers{C. Johnson, R. Khuri and R. Myers, hep-th/9603061.} 

\lr \myk{R.R. Khuri and R.C. Myers, hep-th/9512061.}

\lr \KT{I.R. Klebanov and A.A. Tseytlin, hep-th/9604089.}

\lr \CYY{M. Cveti\v c and D. Youm, unpublished.}
\lr \CS{M. Cveti\v c and  A. Sen, unpublished.}
\lr \AT{ A.A. Tseytlin, hep-th/9604035.}

\lr \green{M.B. Green and M. Gutperle, hep-th/9604091.}
\lr \sen{A. Sen, \mpl  { A10} (1995) 2081, 
 hep-th/9504147. }
\lr \gib{G.W.  Gibbons and K. Maeda, \np {B298} (1988) 741.} 
\lr \dabb{  A. Dabholkar, J.P. Gauntlett, J.A. Harvey and D. Waldram, 
  hep-th/9511053.   }
\lr \CMP{ C.G. Callan, J.M.  Maldacena  and A.W. Peet, 
PUPT-1565,  hep-th/9510134.  } 

\lr \dab{  A. Dabholkar, J.P. Gauntlett, J.A. Harvey and D. Waldram, 
 CALT-68-2028, hep-th/9511053.   }
\lr\susm{J. Maldacena and L. Susskind, hep-th/9604042.}
\lr\DasMat{S. Das and S. Mathur, hep-th/9601152.}

\lr \duf{M.J.  Duff, S. Ferrara, R.R. Khuri and J. Rahmfeld, 
\pl {B356} 
(1995) 479, hep-th/9506057.}

\lr \lars { F. Larsen, private communication.}
\lr \TTT{A.A. Tseytlin,  hep-th/9603099.}
\lr \garf{D. Garfinkle, \pr D46 (1992)  4286.}

\lr \ssen{ 
A. Sen,  \np B388 (1992) 457; D. Waldram, \pr D47 (1993)  2528.} 

\def \A {{\cal A}}
\def \F {{\cal F}}
\lr \dul{M.J.  Duff and J.X. Lu,  \np { B354}  (1991) 141.}

\lr \ttt{A.A. Tseytlin, \pl B363 (1995) 223, hep-th/9509050.}
\lr \TET{A.A. Tseytlin, \pl B251 (1990) 530.}
\lr \lowe{D.A. Lowe and A. Strominger, \prl {73} (1994) 1468, hep-th/9403186.}
\lr \HS {G. Horowitz and A. Strominger, 
hep-th/9602051.}
\lr \HMS {G. Horowitz, J.M. Maldacena  and A. Strominger, 
hep-th/9603109.}

\lr \KT{I.R. Klebanov and A.A. Tseytlin, hep-th/9604166.}
\lr\dull{M.J. Duff and J.X. Lu, \pl B273 (1991) 409. }
\lr\hos{G.T.~Horowitz and A.~Strominger, Nucl. Phys. { B360}
(1991) 197.}
\lr\ght{G.W. Gibbons, G.T. Horowitz and P.K. Townsend, \cqg 12 (1995) 297,
hep-th/9410073.}
\lr\CYYY{M. Cveti\v c and D. Youm, hep-th/9603100.}

\lr\DVV{R. Dijkgraaf, E. Verlinde and H. Verlinde, hep-th/9603126.}
\lr \kallosh {E. Bergshoeff, I. Entrop and R. Kallosh,
 \pr D49 (1994) 6663.}
\lr\suss{L. Susskind, hep-th/9309145.}

\newsec{Introduction}
One of the first systematic attempts to  give a 
statistical interpretation of the  Bekenstein-Hawking (BH) black hole entropy  
 on the basis of string theory was made in \refs{\sen} where 
it was proposed that  entropy 
of extremal electric black holes can be understood in terms 
of counting of the corresponding elementary  supersymmetric (BPS) 
string states. This idea was clarified and  put on a firmer ground 
 in \refs{\CMP,\dab}
by identifying  degenerate extremal electric 
 black hole states with oscillating modes of 
underlying macroscopic string.
While the statistical entropy  of elementary BPS string  states 
is given (in the limit of large charges) 
by 
\eqn\sta{ S_{stat} = 2\pi \sqrt {{c_{eff}\ov 6} N_L}\ ,  \ \ \ \  \ 
N_L =Q_1Q_2 \ , }
(where  we assumed  the simplest case of the
two electric 
charges corresponding to a  
circular dimension   and $c_{eff} =12 $ in type II superstring
and $c_{eff}=24 $ for the heterotic string),
 the semiclassical BH 
entropy of extremal electric black holes vanishes
when computed at the singular $r=0$ horizon.
It was suggested \sen\  that the account of string  $\a'$ 
corrections  may  modify the black hole solution
leading to a new  `stretched' \suss\  position of horizon 
at $r\sim \sqrt{\a'} $
with the entropy $S_{BH} = k \sqrt {Q_1Q_2}$.
The latter has the right  form  to match $S_{stat}$ (see also \peet),  but the 
 coefficient $k$ which should depend on detailed structure 
of $\a'$ corrections in a particular 
string theory (and, indeed,  should be different in
type II and heterotic string theories in order for 
$S_{stat}$ and $S_{BH}$ to have chance to agree)
is hard to  compute  explicitly.

The important  further suggestion  \LW\  was that 
 a precise  test  of the idea  about  the string   
statistical  origin
of the BH entropy   should  be possible in the case of  the $D=4$ 
supersymmetric  dyonic extremal 
  black holes  \refs{\kall,\CY,\US}  for which  $S_{BH}$ is non-zero
already at the semiclassical level\foot{In this Section we shall assume for simplicity that the electric and magnetic charges
 $Q_i$  and $P_i$ are normalised to take  integer 
values at the quantum level. $Q_i$ and $P_i$ in the following Sections 
will differ from integers by  certain factors  discussed below.}
\eqn\ere{
   S_{BH}= 2\pi \sqrt {Q_1Q_2P_1P_2} \ . }
The idea  \LW\ was to interpret the latter  as the statistical 
  entropy  \sta\ of supersymmetric
oscillating  states of a free string  with  tension 
renormalised by the product $P_1P_2$ of  magnetic charges.

The reason for this magnetic renormalisation 
was  explained in \refs{\USS} (see also \TT) 
  starting with  the  conformal \sm 
which describes  the embedding  \US\  of the dyonic
black holes into string theory.  The marginal supersymmetric deformations
of the  conformal \sm  were interpreted  (in the spirit  of \refs{\CMP,\dab})  as describing  degenerate black holes 
with the same asymptotic charges but different short distance 
structure. Since these perturbations  are important  only
at small scales  they can be effectively counted  near 
the horizon ($r=0$). The crucial  observation 
 is that for  $r\to 0$ the  model
 reduces to  a  6-dimensional WZW  theory 
with  level $\k = P_1P_2$ (similar models were discussed in \refs{\chs,\lowe}).

The original suggestions   of \refs{\LW,\USS,\TT,\LWW}  
suffer, however, from the two (related) problems.
First, it appears that if one rescales  the free string oscillator
level $N_L=Q_1Q_2$  by $\k=P_1P_2$ and uses the standard 
expression \sta\ for the statistical entropy 
with $c_{eff}=24$ corresponding to the heterotic string
one  finds the result which is twice the  BH 
expression  \ere. For  type II  theory choice $c_{eff}=12$
one is off by factor of $\sqrt 2$. 
The second problem is why, in fact,  the expected 
  result  should  at all depend  on a choice of  a particular 
string theory in which black hole is assumed to be a solution.
In contrast to the  `quantum' stretched horizon entropy of
 \sen\   the BH entropy \ere\   is  a semiclassical one.  
It  should  depend only on data  contained in the  
 classical solution,   i.e. values and types of asymptotic charges
and the number of   space-time dimensions. 
The entropy \ere\ certainly cannot `feel' the precise 
value  of all internal compact directions of a particular  critical 
string theory. All it may  implicitly `know' about  is
 how 
the four charges are embedded into string theory in a way preserving supersymmetry 
and that relevant microstates should  also be supersymmetric. 

Our aim  below  will be to resolve these problems
and to argue that the approach  initiated in  \refs{\LW,\USS,\TT,\LWW}
does indeed  lead to the quantitative  explanation
of the BH entropy \ere\ and in that sense is consistent with and complementary to the approaches based on D-brane counting
of BPS states (see  \refs{\SV,\CM,\HS,\MV,\mast,\myers,\DasMat,\susm}
and \horr\ for a review). 
 Its advantage is a  direct 
space-time  interpretation   (in particular, a clear identification
of oscillation modes responsible for the BH entropy).

An important  observation
is that the statistical entropy  corresponding to all possible
supersymmetric marginal perturbations of a string theory solution
(non-trivial conformal $\s$-model) 
which represents the dyonic black hole 
is a complicated function of  the charges 
which  reproduces  \ere\ only in the limit when
all charges are large.  Assuming that 
$Q_1,Q_2,P_1,P_2 $ are of order $Q \gg 1$ we may expect to find, symbolically
(ignoring possible $Q^n \ln Q$ terms, etc.)  
\eqn\eem{ S_{stat} = a_1 \sqrt {Q^{4}} +  a_2 \sqrt {Q^3} + 
a_3 \sqrt {Q^2} +  .... + O(Q\inv ) \ .} 
Different types of perturbations may contribute to different 
terms in the exact expression for $S_{stat}$. In the limit when 
magnetic charges  vanish,   the exact expression should reduce 
to the standard  free string BPS entropy  \sta. 
While the $\sqrt {Q^2}$ term in \eem\ should  thus depend 
on embedding into a particular string theory, 
the leading   $\sqrt {Q^4}$ term  should  be universal.
 For this  to  be possible,  this leading term
should  be built out of  contributions  of only certain 
 universal types of  perturbations which are common to 
heterotic and type II embeddings. Other perturbations, in particular,  
 string  oscillations in internal  toroidal   directions 
should   contribute  only to {\it subleading} terms in the entropy.

Indeed,  the relevant perturbations  turn out to correspond
to string oscillations in three  non-compact spatial directions
and one compact direction responsible for  the Kaluza-Klein-type 
embedding  of 
 the two magnetic charges. These dimensions are  indeed `intrinsic' to 
the black hole and do not depend on a choice of a superstring theory.
It is only these {\it four}   directions  that get multiplied 
by $P_1P_2$ in the near-horizon region.\foot{This   
is an obvious consequence  of  the discussion in \refs{\USS,\TT}
but was  overlooked  there: it was expected, following the 
conjecture of \LW, that  all the transverse  string coordinates, 
including the internal toroidal ones 
get magnetically renormalised tension.
The possibility that only perturbations in four of  transverse directions 
are actually renormalised by $P_1P_2$ was  suggested 
to me by F. Larsen \lars.}
In terms of free-oscillator 
description of perturbations 
 (which indeed applies near $r=0$ for large $P_1 \sim P_2$, 
i.e. large  level $P_1P_2$  of  the underlying   WZW model) 
this corresponds to the effective number
of degrees of freedom 
\eqn\rere{ c_{eff}= 4(1 + \ha) = 6 \ , }  
where we have included the  contributions of superpartners 
of the four bosonic  string coordinates. This  result
is the same in type II and heterotic string theory (provided
the black hole solution is embedded into the heterotic 
theory in the  manifestly conformally-invariant 
`symmetric' way). 
The statistical entropy \sta\ 
corresponding to  supersymmetric string oscillations in the
four transverse directions  with the tension or oscillator
 level rescaled
by $P_1P_2$  exactly matches the BH expression \ere.

To make the discussion more transparent,  
in what follows we shall consider in detail   the  
 very similar case of  $D=5$ dyonic  black holes. An
advantage of the present conformal \sm approach  is that 
the treatment of the 
 $D=4$ and $D=5$  dyonic black hole cases is parallel
since both are described by closely related $D=6$ 
supersymmetric conformal $\s$-models \refs{\US,\TT}.
Once the  main conceptual issues are clarified
on the $D=5$  example, generalisation to the $D=4$ case
 is straightforward.

The  conformal model \TT\ describing 
 $D=5$
dyonic black hole with two electric and one magnetic charges
$Q_1,Q_2,P$  can be interpreted as representing
a BPS `bound state' of a fundamental string \duh\  and solitonic 5-brane \refs{\chs,\dul}
or as a `boosted'  $D=6$ dyonic string \duf\ (Section 2).
 The analogs
of the expression \ere\ and \eem\  are 
\eqn\ereu{
   S_{BH}= 2\pi \sqrt {Q_1Q_2 P } \ ,  }
and ($Q_1, Q_2,  P\sim Q \gg 1$) 
\eqn\eemu{ S_{stat} = a_1 \sqrt {Q^{3}} +  a_2 \sqrt {Q^2} 
 +  ...  + O(Q\inv ) \  .  } 
We shall argue  in Section 3 that  the 
supersymmetric marginal perturbations which  reproduce \ereu\
as the leading term in the statistical entropy \eemu\ 
correspond to the transversal  oscillations of $D=6$ dyonic string, 
i.e. oscillations  in the 
four non-compact  directions (spatial dimensions of $D=5$ black hole). Only these  four string coordinates 
  get  their tension effectively 
rescaled by the magnetic charge $P$ (level $\k$) in the near-horizon  (throat) region described by  a WZW theory.
Oscillations in the compact toroidal directions 
(4 in type II and  4+16 in the heterotic theory) contribute only
 to subleading terms in the statistical entropy.
As in the $D=4$ case the effective number of 
degrees of freedom  responsible for \ereu\  is thus $c_{eff}=6$.\foot{Not surprisingly,  the value $c_{eff}= 6$ (same  for $D=4$ 
and $D=5$ black hole cases) appears also in 
related D-brane  (in particular, 
\refs{\susm}) and M-brane  \refs{\DVV,\KT} approaches.
It is interesting to note, however, that the  four relevant 
directions in the present NS-NS description  are orthogonal to 
the solitonic  5-brane, 
while in the R-R (D-brane)   and M-brane descriptions  they are the
tangential R-R  5-brane  directions  which  are  transversal to the R-R string lying within the  5-brane \CM.}

To relate the number of oscillations to 
charges of the classical solution  and to fix the quantisation
of $Q_1,Q_2$ we shall follow   \refs{\dab,\CMP}
and  consider    matching  of    deformed 
  supersymmetric dyonic string  solutions
onto  oscillating string source (Section 4).  Although  trying to  
match onto  a string source may seem surprising in view of 
non-singular nature of the dyonic solution  we shall suggest  that source
interpretation  is  still   consistent  for  solutions
which carry { both}  magnetic  {\it and }  electric  charges
 (sources are not needed only in the case of pure magnetic 
solitons).\foot{This seems
to apply not only to the  $D=6$ dyonic string but also to 
$D=10$ dyonic 3-brane \refs{\hos,\dull}
(even though it is  also  non-singular \ght).}  

The resulting  analogue of the free-string level matching condition 
 makes possible to express the oscillation number 
corresponding to the  four  non-compact transverse directions
in terms of the product of charges $Q_1Q_2P$ and thus to reproduce
the BH entropy \ereu\ as the statistical one \sta.
The generalisation of the discussion 
to the case of $D=4$  dyonic black hole
is straightforward  and will be briefly sketched in Section 5.

\newsec{$D=5$ dyonic black hole and $D=6$ 
dyonic string  behind it}
We shall consider the following $D=5$ extremal 
dyonic  black hole  with the Einstein-frame 
metric \TT\foot{Non-extremal generalisation of this 
solution was  constructed in 
\refs{\CYYY,\HMS}.} 
\eqn\mee{
ds^2_E =- \l^{2} (r)  dt^2 +  \l\inv  (r)  (dr^2 + r^2 d\Omega^2_3) \ , 
}
$$
\l 
 = {r^{2} \ov [(r^2+Q_1)(r^2+Q_2)(r^2+ P)]^{1/3}}\ . $$ 
Supplemented by two electric and one magnetic vector fields and  
scalar fields it  represents an   $N=1$ supersymmetric  solution of extended 
$D=5$  supergravity. We shall assume that all three charges are positive
so that $r=0$ represents a regular horizon.
For $Q_1=Q_2=P$ this metric is equivalent to $D=5$ Reissner-Nordstr\"om one
\refs{\gib,\SV}.
The mass  of this $D=5$ black hole and the Bekenstein-Hawking entropy 
proportional to the 3-area of the   horizon 
are given by ($G_N$ is the 5-dimensional Newton's constant)
\eqn\mass{
 M=  {\pi \ov 4G_N} (Q_1 +Q_2 + P)\  ,  }
\eqn\entr{
{S}_{BH} = 
{{ A} \ov 4G_N} = { \pi^2 \ov 2G_N} \sqrt {Q_1Q_2P}  \ .  }
This black hole background can be embedded into string 
theory in several different ways, depending
on the interpretation (NS-NS or R-R) of the corresponding
charges or vector fields.
Here we shall  follow \TT\ and discuss  the  purely NS-NS embedding
(R-R embeddings were considered in \refs{\SV,\CM}).
The corresponding   background
is an exact  solution in  both   heterotic and  
type II  superstring  theories \refs{\HT,\US}
and may be interpreted as a  BPS `bound state' of  
a closed macroscopic string  \duh\ 
and  a solitonic 5-brane \refs{\chs,\dul}
(5-brane  will be  assumed to be wrapped around 5-torus 
with the string wound around one of its cycles).\foot{Under $SL(2)$
duality of $D=10$ type IIB theory it  becomes 
a solution with RR-charges which can be 
described as  a superposition of a D1-brane  and D5-brane \CM.}
It can be represented by the $D=10$  supersymmetric conformal \sm\
with the following bosonic part 
\refs{\HT,\US,\TT}
  \eqn\lag{
L = (G_{\m\n} +   B_{\m\n })(x)  \del X^\m \bd X^\n  + {\cal R}\P (x) }
$$= 
 F(x)  \del u \left[\bd v +  K(x) \bd u  \right] + 
(g_{mn} +   B_{mn })(x)  \del x^m \bd x^n  + \del y_a \bd y_a 
+  {\cal R}(\p + \ha  \ln F)  \ , $$
where  $u= y_5 -t, \ v=y_5 +t$, 
$x_m$ ($m=1,2,3,4$) are non-compact spatial coordinates, 
 $y_a$ ($a= 1,2,3,4$) and $y_5$  are coordinates of 5-torus, and \chs\ 
\eqn\pep{    g_{mn} = f(x) \delta_{mn}\   ,  \ \ \ \   H^{mnk} = -{\textstyle { 2\ov \sqrt g}}  \epsilon^{mnkp} \del_p \p \ , \ \ \ 
e^{2\p} = f \ , 
\ \ \ \  \del^m\del_m  f =0 \ ,  
}
where $H_{mnk}\equiv 3\del_{[m} B_{nk]}$. 
Note that $\sqrt G e^{-2\P} = \sqrt g e^{-2\p} = f $  and 
$\sqrt g e^{-2\p} g^{mn} = \delta^{mn}$. 
The conformal invariance conditions
reduce to the flat-space ones
\eqn\harm{ \del^m\del_m  F\inv =0\ , \ \ \ \   \del^m\del_m K=0\ , }
 i.e.
the model is parametrised by the  three harmonic functions $f,F\inv,K$
depending on four non-compact 
transverse coordinates $x^m$.
This is a reflection of the BPS-saturated  nature of the  solution.
The special one-center  choice  ($r^2\equiv x_m x_m $)\foot{We 
change notation slightly compared to \TT:  in \TT\ $K$ 
had  asymptotic value  1 and at the same time $v$ was $2t$, not 
$v=y+t$ as here. We also 
flip  the definitions of  $Q_1$ and $Q_2$.}
\eqn\choi{
f= 1 +{P\ov r^2} \ , \ \ \ \ \ 
F\inv = 1 +{Q_1\ov r^2}\  , \ \ \ \ \ 
K={Q_2\ov r^2} \ , }
leads upon compactification along $y_1,...,y_5$ 
to the $D=5$ black hole \mee.
The corresponding $D=10$ dilaton and `radius' of $y_5$ 
\eqn\dia{
e^{2\P} = Ff= {{ r^2 + P}\over{r^2 + Q_1}}\ , \ \ \ \ \ 
 e^{2\s} = F(1+K)= {{ r^2 + Q_2}\over{r^2 + Q_1}} \ , } 
are regular both at $r=0$ and $r=\infty$.

Setting  $Q_2=0$ and omitting the trivial 4-torus part
 the model \lag\  becomes 
\eqn\qeq{ 
L  =
 F(x)  \del u\bd v  
+ f(x)\del x^m \bd x_m   
      + B_{mn}(x) \del x^m \bd x^n  
+  \ha {\cal R} \ln [F(x)f(x) ] \ ,  }
and may be also interpreted  as describing  the dyonic  $D=6$ 
string \duf: $B_{\m\n}$ has both electric  $Q_1$ 
and magnetic $P$ charges.  Switching on  $Q_2$ corresponds 
effectively to adding  momentum along the string direction. 
 One can also 
generalise this  conformal model by introducing  a rotational 
parameter \TT\ (see  also  Section 3).

The non-singular solitonic nature of the 5-brane  model \chs\ 
is responsible for the regularity  of the  dyonic   theory  at $r = 0$.
Here the throat limit $r\to 0$ is described
by  the  direct product of    $SL(2,R)$ ($u,v,\r$) 
and $SU(2)$ ($\t, \vp, \psi$) 
WZW models with equal levels proportional to $P$ \foot{The `transverse' ($\r, \t, \vp, \psi$) part
 of the  throat region  model  is exactly the  same as in the  5-brane case 
 \chs\  (where $Q_1=Q_2=0$) except for the fact that 
here the dilaton $\P$ 
is { constant}
in the $r\to 0$  region, i.e. the string coupling is not blowing up.
The angular  coordinates $\t, \vp, \psi$
(appearing in $dx^mdx_m = dr^2 + r^2(d\t^2 + \sin^2\theta d \vp^2 
+ \cos^2\t d \psi^2)$)   are  related to the 
  Euler angles of $SU(2)$  
($0 \leq \t' \leq \pi, \   0 \leq \vp'
 \leq 2\pi,\ 
0 \leq \psi' \leq 4\pi$) by
$\t= \ha \t', \   \vp= \ha (\vp' + \psi'), \     \psi=  \ha (\psi' - \vp')$.} 
\eqn\tet{
 I  =  {1 \ov \pi\a' }\int d^2 \s  L_{r \to 0} 
=  {1 \ov \pi\a' }\int d^2 \s  \left( e^{-2\r  }   \del   u \bd  v  + Q_2  Q_1\inv \del   u  \bd   u  \right) 
}
$$
  + \ 
{\k \ov \pi }\int d^2 \s  \left[\del \r \bd \r
+ \del \t \bd \t  +
 \sin^2\theta \del \vp \bd \vp   + \cos^2\t \del \psi \bd \psi
+  \ha  \cos 2 \t (\del \vp \bd  \psi - 
\bd \vp \del \psi)   \right]  ,  $$
\eqn\uuu{
  \r \equiv  \ln {\sqrt{Q_1 } \ov  r} \   \to \infty \ , \ \ 
\ \ \ \ \ \  P\equiv  {\a'} \k  \ . }
We  have omitted the constant dilaton term. The $(1,1)$ supersymmetric version 
of this  model has free-theory central charge: 
$c= [3(1 + {2\ov \k}) + {3\ov 2}] + [3(1 - {2\ov \k}) + {3\ov 2}] = 6(1 + \ha)$.
Since $P$ is related to the coefficient of a Wess-Zumino term in the string 
action, i.e. is proportional to the integer level $\k$, it is 
 quantised in  units of  $\a'$. The quantisation of the
 electric charge 
$Q_1$  can then be deduced  from the  Dirac condition  applied to 
the antisymmetric tensor in  $D=6$. 
The   quantisation of  $Q_2$ 
is the same as in the fundamental string case  \refs{\duh,\dab}
and will be discussed  in Section 4  by matching on a string source
($Q_1$ is then related to the winding number and $Q_2$ 
to the momentum of a string source). 

\newsec{Oscillating $D=6$ dyonic string states  as degenerate  dyonic 
$D=5$  black holes}
Our aim  will be  to follow 
the previous suggestions \refs{\CMP,\dab,\LW,\USS}
and to try to reproduce  the BH entropy \entr\
as a statistical entropy 
related to existence of an infinite  family of  more general 
$D=5$ black hole solutions which asymptotically look the same
as \mee, i.e. have the same charges, but differ at scales
of order of compactification scale 
 (equal to the radius of the string direction $y_5$).

\subsec{Deformed sigma model  and conformal invariance conditions}

Demanding   preservation of
the same amount of  supersymmetry, i.e.  the   BPS property 
(or related property of manifest  exactness of the solution),  
these more general backgrounds are described  by 
the following extension  of \lag\ \refs{\HT,\USS}\foot{There are other marginal deformations
of the chiral null model which  break supersymmetry \TTT.
These are expected to suffer from string $\a'$ (and loop) corrections
so that  their contribution to the entropy is hard to determine
precisely (presumably they give subleading contributions 
to  statistical entropy in the limit when all three charges are large).
It is  also likely that these deformations are unstable  and 
 cannot be interpreted as true microscopic black hole states.}
\eqn\yyy{
   L =  F(x)  \del u \left[\bd v + K(u,x) \bd u 
 +   2{\cal A}_n (u,x)  \bd  x^n 
+ 2\A_a (u,x)  \bd  y^a   \right] }
$$  +  \ 
(g_{mn} +   B_{mn })(x)  \del x^m \bd x^n  + \del y_a \bd y_a 
+  {\cal R}(\p + \ha  \ln F)  \ , $$
where   $\A_m$ and $\A_a$ correspond to `deformations'
in four non-compact $x^m$ and  compact $y^a$ directions respectively. 
The requirement of supersymmetry is directly related to 
the  chiral null structure of \yyy\ \refs{\HT,\kallosh}, 
i.e. to the dependence of deformation functions 
on only {\it one}  null (string-like) 
direction.

The conditions of exact conformal invariance of this model \refs{\USS,\TT}
are again $\del^m\del_m F\inv=0$
and\foot{
If $x^i$ denotes the set of all transverse directions (both compact and non-compact)
 the condition of marginality of the perturbation
$\A_i (u,x) \del x^i  $ (i.e. $ {\cal A}_n (u,x)  \bd  x^n +   \A_a (u,x) \bd y^a$
in the present case) is, in general, 
$   \nabla_{+ i }(e^{-2\p} {\cal F}^{ij} ) = 0$ \ 
($\Gamma^i_{+jk}
=  \Gamma^i_{jk} + {1\over 2} H^i_{\ jk}$),  \   
i.e. $\del_i   (e^{-2\p} \sqrt g {\cal F}^{ij})  -
\ha e^{-2\p} {\sqrt g} H^{kij} {\cal F}_{ki}
=0 $, where $g_{ij}$, $B_{ij}$ and $\p$ are the couplings
of the  transverse conformal theory. Since the latter is trivial
in $y_a$-directions and  $\A_a$ does not depend on $y_a$
one concludes that  $\A_a$  should satisfy the free Laplace 
 equation in 4 non-compact directions.} 
\eqn\kkk{ \nabla^m (e^{-2\p} \del_m K)  
 -2 \del_u \nabla^m (e^{-2\p} \A_m)=0 \ ,  \ \ \  i.e. \ \ \ 
 \del^m(\del_m K - 2\del_u \A_m) =0 \ ,  } 
\eqn\aap{ \del^m\del_m \A_a =0 \ , \ }
\eqn\ssp{ 
 \nabla_m   ( e^{-2\p}  {\cal F}^{mn})  -  e^{-2\p}  H^{nkl} {\cal F}_{kl}
=0  \  ,    }
where $\F_{mn } \equiv  \del_m \A_n - \del_n \A_m$.
This  system is invariant under $K \to K + \del_u \s(u,x), \ 
\A_m \to \A_m + \ha \del_m \s(u,x)$ induced by $v\to v + \s(u,x)$
in  \yyy\ (the $u$-dependent part of  $K$  can be absorbed into 
 ${\cal A}_n$ by a  redefinition of  $v$). 
One may  assume, for example,  that 
$\del_u \del^m\A_m=0$ as a gauge choice so that the equation for $K$ 
becomes again 
the flat Laplace one   $ \del^m\del_m K =0$.\foot{Alternatively, one may define $\A_m'=\A_m - \ha  \del_m \int du  K (u,x)$
and  integrate \kkk\ to get  $ \del^m\A_m' = h(x)$.}

The equation   for $\A_m$ \ssp\
can be put in the following simple form  \TT
\eqn\yuy{    
\del_m   (e^{-2\p}  \sqrt g  {\cal F}^{mn}_+ ) =0   \ ,
\ \ \ i.e.\ \  \ \ \del^m   (f\inv    {\cal F}_{+ mn} ) =0 \ , 
}
\eqn\popt{
 \F_{+mn} \equiv  {\cal F}_{mn} + {\textstyle{1 \ov 2 \sqrt g }}  {g_{mp}g_{nq} }   \ep^{pqkl} {\cal F}_{kl}  =    {\cal F}_{mn} + {\ha   } \ep^{mnkl} {\cal F}_{kl}         \ .    }

\subsec{Perturbations in compact and non-compact directions and entropy }
There are several important conclusions which follow from these 
exact marginality conditions.
First, they impose essentially 
 no restrictions on $u$-dependence
of $K,\A_a,\A_m$,  i.e.  as in the case of the `free'  fundamental string 
\refs{\garf,\ssen, \HT,\CMP,\dab}  they represent   various 
`left-moving' waves propagating along the string or 
oscillating string modes.
  Assuming  Kaluza-Klein compactification
along  the string direction $y_5$,   the \sm \yyy\ 
describes  a family 
of  BPS saturated $D=5$ black hole  backgrounds which  are  the same
as \mee\ at distances larger that the radius of $y_5$ but 
are  different at smaller scales \refs{\CMP,\dab,\LW}. 

  Second,  deformations in the compact $y_a$ directions
described  by $\A_a$ (charge  waves along the string)
satisfy the free Laplace equation and thus 
are effectively decoupled from the non-trivial non-compact 
part of the model, i.e. do not depend on  the `magnetic'   function $f$.
At the same time, the string oscillations  in {\it non-compact}
 directions 
described by $\A_m$ (in particular, rotational 
perturbations) are non-trivially coupled to the 
function $f$, i.e.  are, in general,  sensitive to the value 
of the magnetic charge $P$.

The fact that perturbations in compact 
directions do not `feel'   $P$
 suggests that  counting of the corresponding 
oscillating string states 
should  give  exactly the same  result as in 
the case of the  `free' fundamental string, 
i.e. the associated statistical entropy   should 
have nothing to do with the  leading-order semiclassical 
BH  entropy \entr. Indeed, it turns out that the matching
condition relating `compact'  oscillation number to charges 
has the form $N_L \sim Q_1Q_2$, i.e. also does not involve $P$.

This is a natural conclusion: the entropy  related to 
oscillations in  $D_{int}$ compact toroidal directions 
 ($\sim \sqrt { {c_{eff}\ov 6} 
Q_1Q_2}$, \ $c_{eff} = D_{int}(1 + \ha)$) 
 depends on their
number,  i.e. on a   particular string theory 
($D_{int}= 4$ in 
 type II theory 
and $D_{int}= 4 + 16=20 $ in  heterotic theory)
 while the black hole solution \mee\ and its semiclassical BH entropy 
\entr\ are certainly universal. 
These oscillations  do contribute to the entropy (as in the
purely electric extremal black hole case \refs{\sen,\peet}), but they 
produce  a {\it subleading } contribution  when  all three  charges 
 $Q_1,Q_2$ and $P$ are large and  of the same order (cf. \eemu). 
The  $P=0$ limit  of the exact expression for the statistical BPS entropy 
is expected to  reproduce  the  
entropy associated with the stretched horizon of the 
extremal electric black hole \sen\ 
with   the coefficient  which is 
 different in heterotic and type II theories
(since the structure of $\a'$ corrections  and thus the position 
of the stretched horizon  is different in the two theories).

At the same time, the  counting of perturbations
in {\it non-compact}  dimensions should depend only 
on their number (four) and the values of all {\it three}  charges --  
 the data intrinsic to $D=5$ black hole, i.e.
independent of   embedding into  a particular string theory  (assuming only 
that supersymmetry is maintained). 
The corresponding statistical entropy 
 should thus be   universal as is    the BH entropy \entr.
Below we shall  confirm that the two entropies indeed match.
The conclusion is thus  that  only the supersymmetric transverse  
perturbations of the dyonic $D=6$  string 
are responsible for the semiclassical BH entropy \entr.

\subsec{Examples of conformal deformations}
The equation \aap\ is solved by
\eqn\cha{ \A_a (u,x) = {q_a(u) \ov r^2}  \ .}
Since $u=y_5-t$ and $y_5 \equiv y_5 + 2\pi R$, 
the   `charge wave'  functions $ q_a(u)$ are periodic, i.e.
are  given by  Fourier series expansions, 
$ q_a(u) = \bar q_{ a} + \td q_a(u),  \ 
\td q_a(u) =  b_a \ \sin (R\inv u)  + ... $.
If the   constant parts $\bar q_{ a}$ are non-vanishing, they 
produce  additional  asymptotic 
(`left') electric charges of the $D=6$ string 
and thus of the  corresponding more general 
$D=5$ extremal black hole. The  BH entropy of the latter 
can be obtained  by making  the shift 
 \eqn\yty{ Q_1 Q_2 \rightarrow  Q_1 Q_2 - \bar q_a^2 \  } 
of the product of the two electric charges   in  \entr\    (see  also  Section 4).

A particular solution of \yuy\ is  found by 
imposing the  flat-space anti-selfduality condition 
\eqn\self{ {\cal F}_{+mn} =0 \ . }  
 In terms of
 the angular coordinates $(\t$,$\vp$,$\psi)$  in \tet\  
which are related to $x^m$ by 
$$ x^1+ix^2 = r\  \sin\ \theta\ e^{ i \vp }, \ \ \   
 x^3 + ix^4=  r \ \cos\ \theta\  e^{ i \psi }, \  $$
 $$ dx^m dx_m  =
 dr^2 + r^2 (d\theta^2 +
 \sin^2\theta d\vp^2  + \cos^2\t d\psi^2)\ , $$
we find that \yuy,\self\  are satisfied  by  \TT\
\eqn\selo{
\A_m (u,x)  dx^m =  { \g (u) \ov r^2} ( \sin^2 \t d\vp + \cos^2 \t d\psi)
\ .  } 
Here $\g(u)$ is an arbitrary periodic  function, 
 $ \g (u) = \bar \g  + \td \g (u), \  \td \g (u) = b \  \sin (R\inv u) + ... $.
The constant part $\bar \g$  has the meaning of a rotational
parameter. In fact, dimensionally reducing along $y_5$ (i.e. averaging over  $u$)
one finds the following modification of the $D=5$ metric \mee\ 
\eqn\ere{
 ds^2_E =- \l^{2} (r)  (dt +  \bar \A_m   dx^m)^2 + \l\inv  (r)  dx^m dx_m\ ,  }
which is  the  supersymmetric 
rotating  generalisation of the 3-charge $(Q_1,Q_2,P)$  extreme dyonic $D=5$ black hole \TT\ (the special case $Q_1=Q_2=P$ of this solution   
 was found in \MV).
 In addition to the  mass 
(still given by \mass)  and three charges there are also 
 two equal angular momenta in the two  orthogonal planes
 \eqn\jjj{ J_\vp =  J_\psi\equiv  J=  {\pi \ov 4G_N} \bar \g \ . }
The corresponding BH entropy  is found to be given by 
\entr\ with 
  the product of all {\it three}  charges   shifted by $\bar \g^2$
 \eqn\ytyh{ Q_1 Q_2 P  \rightarrow  Q_1 Q_2 P  - \bar \g^2  \ .  } 
 Comparison of \yty\ and \ytyh\ 
illustrates  the difference between parameters  associated with 
supersymmetric marginal deformations  in  internal 
and non-compact directions. 

On may assume for simplicity that 
the mean values of the  above deformations vanish ($\bar q_a=0, \ \bar \g=0$,  etc.),  i.e. that 
they do not introduce new asymptotic parameters  so that   
the  BH entropy is given by the original expression \entr.

\subsec{Perturbations in the throat region}
Since we would like to  count such 
non-compact  deformations $\A_m (u,x)$
which decay at large $r$  this can be done 
 near $r=0$, i.e. at the `throat' (or  $D=5$ horizon).
Indeed, it is  in this region that 
the degeneracy between different members 
of the family of $D=5$ black holes with the same asymptotic charges 
is lifted. 

This is an important simplification  since 
the $r\to 0$ limit of the basic \sm   \lag\
is given by the  WZW theory \tet.
The problem can then be reformulated in terms of counting 
of  supersymmetric 
  (chiral null) marginal  deformations of the model \tet\
\eqn\ypo{
I' = I  + {1 \ov \pi\a' }\int d^2 \s  \bigg(F(x)[K'(u,x) \del u \bd u +    2 \A_m (u,x) \del u \bd x^m ]\bigg)_{r\to 0}  } 
$$  = \ 
 I  + {1 \ov \pi\a' }\int d^2 \s\   Q_1\inv  [  k(u) \del u \bd u
+
2 a(u)  \del u \bd \r
  + 2 a_s (u, \b)  \del u \bd \b^s]     \ ,    $$
where $\b_s$ ($s=1,2,3$) denote  the three angular coordinates 
$\t,\vp,\psi$. The
  contribution of $u$-dependent part $K'(u,x)$ 
of $K(u,x)$  can be omitted without loss of generality. 
Indeed, if  $K'(u,x) = h_m (u) x^m  + {  \td Q_2(u)\ov r^2} , $\ 
$\td Q_2(u)= Q_2(u)- Q_2$,  
then  (in contrast to the fundamental string case \dab) 
the oscillation part $h_m (u) x^m$ of $K$ drops out 
 while $k(u) = \td Q_2(u)$  has zero mean value (see also below).

Another crucial point is that we are actually  interested in 
the limit of large charges. 
For  large $P$, i.e.  for large level $\k$ of  WZW model,
the count of perturbations should be the same as in the  
 theory  of four {\it free}   fields  $\r,\b_s$
 (interactions between the fields in WZW action in \tet\ are suppressed by $1/\k$).\foot{Note that the time-like $SL(2,R)$ part of the model \tet\ 
does not enter the discussion in any essential way.}  
The only difference  as compared 
to the free string case will be  due
to the fact that the kinetic terms 
of the four  fields $\r,\b_s$ have the factor $P$ 
while such factor is absent in front of the perturbations in \ypo.
As a  result, there  will be an extra factor of $P\inv $ in the relation
between the number of `left-moving' oscillations  $N_L$ and the
product of electric charges $Q_1Q_2$, 
i.e. $ Q_1Q_2 \sim P\inv N_L$.
Indeed, integrating out the four transverse fields in the limit of large $P$
we find from \tet, \ypo
\eqn\stell
{  I'=  {1 \ov \pi\a' }\int d^2 \s \left[ e^{-2\r  }   \del   u \bd  v 
 +  E (u)   \del   u  \bd   u   \right] \ , }
\eqn\trrr{ 
 E (u) =  Q_1\inv Q_2 +  Q_1\inv  k(u)
  -   P\inv Q_1^{-2} \g^2_m(u)   + O(P^{-2}) \   }
Here $\g_m=(\g_s,\g_4)$,\  $\g_s(u) $  is the  linearized  (`free-theory')  part of $a_s(u,\b)$ and
$\g_4 (u)  \equiv a(u)$.  
As we shall find  in  Section 4, the analogue of the level matching 
condition for the free fundamental string \refs{\CMP,\dab}
which relates the oscillation level numbers to the charges of the background 
 solution  is 
\eqn\lvq{
\overline{ E (u)} =0 \ , \ \ \ \  \ \ \ 
 \overline{  E (u) }  \equiv {1 \ov 2\pi R} \int^{2\pi R}_0 du \   E (u) \ ,  }
i.e., to the leading order in $1/P$,  
\eqn\trew{
Q_1Q_2 P = \overline{ \g_m^2(u)}  \ . }
This is to be compared with the condition one 
finds after adding the perturbations \cha\ 
in  the compact 
internal directions  and integrating over  $y_a$  
\eqn\intr{
I''=  {1 \ov \pi\a' }\int d^2 \s \big( \del y_a \bd y_a + 2[F(x)\A_a(u,x)]_{r\to 0} \del u \bd y^a + ...\big)
} $$ =  \  {1 \ov \pi\a' }\int d^2 \s [ \del y_a \bd y_a + 2Q_1\inv  q_a(u) \del u \bd y^a + ...] \  \rightarrow  {1 \ov \pi\a' }\int d^2 \s [ E(u) \del  u \bd u + ... ] \ ,  $$ 
\eqn\tewq{
E(u) =  Q_1\inv Q_2 - Q_1^{-2} q^2_a(u) + ... \ . }
Then  \lvq\  leads to  the relation 
\eqn\treq{
Q_1Q_2  =   \overline{q^2_a(u)}   \ ,   }
implying that the  contribution of   $q_a$-oscillations to statistical entropy 
compared to that of $ \g_m$-oscillations 
is subleading in the limit of large $P$.

The above discussion can be illustrated on the example 
of the special 
rotational  perturbation \selo\  which  corresponds to  the chiral $SU(2)$ 
Cartan current deformation  
\eqn\rer{ I' = I  + 
{1 \ov \pi\a' }\int d^2 \s  \   Q_1\inv  \g (u) \ \del u \ \bar J_3 \ , }
$$
\bar J_3= 2({\rm \sin}^2 \t \bd \vp + {\rm cos}^2 \t \bd \psi) \ . $$ 
It is possible to decouple $u$ from $SU(2)$ angles 
(or, equivalently,  to integrate the latter out) for generic $P$
by making  a field redefinition.
Introducing 
\eqn\reddd{ \td \vp = \vp +  P\inv Q_1\inv \hat \g (u) \ , 
\ \ \ \ \td \psi  = \psi +  P\inv Q_1\inv \hat \g (u) \ ,
 \ \ \ \ 
 \hat \g (u) 
\equiv  \int^u_0  du' \g (u') \ ,   }
we can represent the  action in the form similar to \tet\  
\eqn\ter{ 
 I'
=  {1 \ov \pi\a' }\int d^2 \s \left[ e^{-2\r  }   \del   u \bd  v 
 +  E (u)   \del   u  \bd   u  \right] 
}
$$
  + 
{P \ov \pi \a' }\int d^2 \s 
 [\del \r \bd \r
+ \del \t \bd \t  +
 \sin^2\theta \del \td\vp \bd \td\vp   + \cos^2\t \del \td\psi \bd \td\psi
+  \ha  \cos 2 \t (\del\td \vp \bd \td \psi - 
\bd \td\vp \del \td\psi)],  $$
\eqn\rerw{  E (u) =   Q_1\inv Q_2 - P\inv Q^{-2}_1  \g^2 (u) =
  P\inv Q^{-2}   [ Q_1 Q_2 P - \g^2 (u) ] \ .  }
In this  special case  the level matching condition \lvq\ is  
(cf. \ytyh)
\eqn\pop{ Q_1 Q_2 P =  \overline{ \g^2 (u)} = 
\bar \g^2 +  \overline{ \td \g^2 (u)} \ . } 
When $\g(u)$ is assumed to contain 
a constant part (i.e. $\hat  \g (2\pi Rn) = 2\pi R \bar \g \not=0$)
 one finds that the corresponding rotational parameter
should be quantised  \TT:
demanding  that $\td \vp, \td \psi$  should have  the same  $2\pi$
periodicity as $\vp, \psi$   we  get    
$  \bar \g  R P\inv Q_1\inv  = l=$integer.
Then the quantisation of $Q_1= {4 G_N \ov \pi}  {wR\ov  \a'}$
 (see Section 4)
and  $P= \k \a'$ 
leads to  the quantization of $ \bar \g= R\inv PQ_1  l$,   i.e.  to the conclusion that the angular momentum $J= {\pi \ov 4 G_N}\bar \g $  should take integer values, 
$J= \k w l=n$.

\newsec{String sources and level matching condition}
To establish a  relation
between  the charges (parameters of the classical solution)
and the   oscillator level  numbers 
(and also to  determine  the quantisation condition for $Q_1$ and 
$Q_2$)   we shall   consider 
the analogue of the procedure of matching on a string source
used  in  \refs{\duh,\dab}.

The backgrounds corresponding to the 
 $D=5$ dyonic black hole and $D=6$ dyonic string model \lag\  
are non-singular at $r=0$. 
Contrary to what one might expect, we believe
  that  it is  consistent to assume 
that having both electric and magnetic aspects,  
the solitonic $D=6$ string, like purely `electric' 
 fundamental string, still needs  to be supported by a source 
at the origin.
This  interpretation
is  important  in order to  be able to  relate 
the macroscopic charges to microscopic 
string oscillations.

A  systematic approach  is  
to   start with equations  always containing
source terms and   to see whether  the 
sources contribute or not for specific 
choices of backgrounds. This  makes possible to  discuss both 
`solitonic' (source-free)   and `elementary' 
(supported by sources)  solutions   as well as 
 intermediate `dyonic' ones 
from a  unified  point of view. 

\subsec{Fundamental string }

Let us first review  the source interpretation 
of the fundamental string solution \duh. While the equations of conformal invariance of the
\sm corresponding to the fundamental string (i.e. \lag\ with $F\not=1,\  K=0,\ f=1$)
are formally satisfied without need to introduce a source
at the origin ($R_{\m\n} + ...=0$ implies
$F^2 \del^m\del_m F\inv=0$ which is satisfied at all points, see  \ttt)
there is  also an alternative   `string source' interpretation
\refs{\duh,\dab}  consistent with  singularity of the 
background at the origin. 
One starts 
with the `combined' action
 \eqn\acti{
 -{1\ov 16\pi G_N^{(D)}}\int d^D x \sqrt G e^{-2\P} ( R + ... ) + {1 \ov 4\pi \a'} \int d^2 \s \sqrt g g^{pq}  G_{\m\n} (x) \del_p x^\m \del_q x^\n + ...\ ,  }
  and  thus obtains  the set of equations
containing (see \dab\ for details)\foot{The dilaton (or conformal anomaly)  equation
$ R - { 1\ov 12}  (H_{\m\n\l})^2 
- 4 (\del_\m \P)^2 + 4 \nabla^2 \P =0$ 
 is not modified  since the  (classical) string source is not coupled to the dilaton. This  implies that the $O(G^{\m\n})$  term  in the above equation 
coming from the  variation of $\sqrt G$ vanishes separately.}
\eqn\eqr{ \sqrt{G } e^{-2\P} (R^{\m\n}  + ...) =  {4G_N^{(D)} \ov \a'}  \int d^2 \s  \  \del_p x^\m_0 \del^p x ^\n_0 \ \d^{(D)} ( x-x_0 (\s)) \ . }
In the simplest case of flat transverse space
and $K=0$ one finds  that the   classical string 
source equations  and  the condition of 
vanishing of the 2-d stress tensor (following  from variation over
 2-d metric $g_{pq}$ which is 
 set equal to $\delta_{pq}$ in \eqr)
 are satisfied by the static winding string configuration $x_0^\m$:
$u_0 = 2R w \s_+, \ v_0= 2Rw \s_-, \ x^i_0 =0$,\  where $w=\pm 1, \pm 2, ...$
and 
$\s_\pm \equiv  \s \pm \tau$, $\  0< \s \leq \pi$.  
Then \eqr\ implies
that  (note that here $\sqrt G e^{-2\P}=1$)
\eqn\uuue{  \del^m\del_m F\inv = -\m \d^{(D-2)} (x) \  , } $$ 
\m =   - {8G_N^{(D)} \ov \a'}  \int d^2 \s   
 \del_+ u_0 \del_- v_0 \delta(u_0) \delta(v_0)  = { 8G^{(D)}_N \ov \a' } w \ , $$ 
so that in the  1-center case ($G^{(D-1)}_N  = G^{(D)}_N/2\pi R$)
\eqn\tetr{
F\inv = 1 + {Q_1\ov r^{D-4}}\ , \ \ \ \ \ \ 
Q_1=  {16\pi  G^{(D-1)}_N \ov  (D-4) \om_{D-3}}\cdot {wR \ov \a'}\ , }
i.e.  for the   $D=6$ string ($G_N \equiv  G^{(5)}_N$)
\eqn\nor{
Q_1=  {4 G_N \ov \pi }\cdot {wR \ov \a'}\ .  }
The $vv$-component of \eqr\ gives 
the equation for $K$  in the form 
\eqn\kkik{ K \del^m\del_m  F\inv + F\inv \del^m\del_m K = 
- {8G_N^{(D)} \ov \a'}  \int d^2 \s   
 \del_+ v_0 \del_- v_0 \delta(u_0) \delta(v_0) \d^{(D-2)} (x) \ . }
The solution   $K= Q_2/r^{D-4}$ of $\del^2 K=0$ 
cannot be matched 
onto string source  \dab.
If one requires that all solutions should be supported by sources 
one thus arrives at the condition\foot{Analogous condition was 
suggested in \CMP\ from  the requirement that the singularity at $r=0$ 
should 
be  null.}
\eqn\tet{ E(u) \equiv [F(x) K(u,x)]_{x\to 0} =0 \ . }
A   point of view which we shall adopt here  
is to admit the possibility that the `basic' macroscopic 
solution  may still be described  by  $K= Q_2/r^{D-4}$.
Even though this  term cannot be directly   matched onto  source
it may  be considered  as an effective description of momentum 
flow since  there is a related  oscillating solution 
 that gives the same space-time background after averaging
over compact direction \dab.\foot{The oscillating fundamental
 string can be described either by 
 $\A_m = g_m(u) $,  or equivalently, 
by  $K= h_m(u) x^m$  (both choices  are  obvious solutions of 
conformal invariance equations and are related by a redefinition of $v$), where $g_m$ and $h_m$
are related to the profile functions $f_m(u)$ of oscillating string,
$g_m =f'_m(u) , \ h_m = 2f''_m(u)$ \refs{\CMP,\dab}.} 
$Q_2$   can then be identified  with  momentum
$p = m/R$ along the string direction (this  interpretation  is also  consistent with $T$-duality in $y_5$ direction 
which interchanges $F(x)$ and $K(x)$  
 and  thus $Q_1$ and $Q_2$  \HT\ as well as the winding 
and momentum numbers). 
In $D=6$  one finds \dab\
\eqn\mou{
Q_2=  {4 G_N \ov \pi } \cdot {m \ov R}\ .  }
 The `matching on source' condition
may  be imposed only on  oscillating generalisations of the basic background.
The minimal   requirement is   that \tet\ 
should  be satisfied  in the average sense   
\eqn\teti{  \int^{2\pi R}_0 du \ E(u)  =0 \ . }
This condition can  be also interpreted 
as the standard classical  Virasoro level matching  condition  $L_0-\bar L_0=0$
or $ \int^\pi_0  d\s (T_{++} -T_{--}) =0$ 
applied to   the solitonic string  described by   (collective-coordinate) \sm 
$L= F(x) \del_+ u[\del_- v + K(u,x) \del_- u] + \del_+ x_m \del_- x_m$
for the basic static winding   source 
configuration $u_0 =2Rw \s_-,\  v_0 = 2Rw \s_+,\  x_0^m=0$. 

\subsec{Dyonic string}
In the case of the 
5-brane or  solitonic $D=6$ `magnetic' string  
 described   by \lag\ with 
$F=1,\ K=0, \ f\not=1$, 
 the equations with source terms
which follow from 
\acti\  are satisfied  with the  left and right parts
of \eqr\ being separately zero.  This is in 
agreement with  expectation that a solitonic solution
need not be supported by a source. For example, the l.h.s. part of  $(mn)$
component of \eqr\
reduces to (see \pep) 
\eqn\rerq{
 f^{-1 }  \Delta_\bot f  \ , \ \ \ \ \ \ 
\Delta_\bot \equiv  {1 \ov \sqrt g  e^{-2\p}} \del_m ( \sqrt g  e^{-2\p} g^{mn} \del_n ) = f\inv \del^m\del_m  \ , }
and thus vanishes automatically for the harmonic function $f$
(the same is true also if we start with \eqr\ divided by 
$\sqrt G e^{-2\P}$). 

When both $F$ and $f$ are assumed to be non-trivial, 
i.e. for a  `bound state' of  fundamental string and 5-brane 
or the $D=6$ dyonic string \qeq,   the equation for $f$ 
is still satisfied automatically while the  equation for $F$
($(uv)$ component of \eqr, cf.\uuue)
\eqn\rery{ f \Delta_\bot F\inv  = 
\del^m\del_m F\inv = -\m \d^{(4)} (x) \ ,   }
 does {\it not}  change its form compared to 
the free fundamental string case
and thus  still needs a source for its support. 
Note that if we  started with  \eqr\  written
in a different  form  with  both parts  divided 
by $\sqrt G e^{-2\P}$ (here equal to $f$)
the conclusion would be different:
we would  get 
\eqn\uuuy{\Delta_\bot F\inv  = 
f\inv \del^m\del_m F\inv = -\m f\inv \d^{(4)} (x) \ , } 
so that the left and right parts would  vanish separately
for  harmonic $f$ and $F$ in \choi.
It seems, however,  that  it is \eqr\ (and thus \rery)  
that is the consistent form  
of the  equation  with source  since 
it directly follows from \acti.\foot{This  action 
 may be understood  as resulting from a 
resummation of string loop corrections  \TET\
which  implies   that there should be no extra 
factors in front of the string metric in the source term.}

Once  the  source interpretation 
of the dyonic string solution is accepted, 
one is able (as in the free fundamental string case) 
to relate $Q_1$  in $F$  to 
the string source winding number $w$  and  $Q_2$ in $K$ 
to the  quantised momentum number $m$, i.e.
to  get again the expressions for the  quantised charges 
 \nor,\mou.
One  can also    obtain again the 
level matching condition \teti, i.e.
the  condition of the vanishing of the mean 
value of the total coefficient $  E (u)$ of the  $du^2$ term in the metric 
(put into the form without off-diagonal $du$-terms) at the core.
This condition \lvq\  was already  applied 
 to perturbations in the near-horizon  region in  Section 3.4.

\newsec{Statistical entropy}
As was already  mentioned above, 
for the purpose of counting perturbations relevant 
for reproducing  the leading term in the black-hole  entropy 
it is sufficient to consider  the region near 
 the origin $r=0$.
Returning to the discussion  of the throat region 
perturbations  in Section 4.1  let us  
express the level matching condition  in terms of the 
quantised charges. 
Using the relations between 
the charges  and  integer quantum numbers
\nor,\mou,\uuu\ we thus find for the `compact' and `non-compact'
 perturbations (see \treq,\trew)
\eqn\yey{ wm =N_L^{(q)} \ , \ \ \ \ \ \  N_L^{(q)} = {\pi^2 \a' \ov 16 G^2_N } \overline{q_a^2 (u) } \ , }
\eqn\yiy { wm\k =N_L^{(\g)}\ , \ \ \ \ \  
N_{L}^{(\g)} = {\pi^2 \ov 16 G^2_N }  \overline{\g^2_m (u) } \ . }
The normalisation coefficient 
 ${\pi^2 \ov 16 G^2_N } $ relating $ \overline{\g^2_m  (u) }$ 
to  integer oscillator number can be checked  by considering   
the special case of  rotational perturbation
\rer,\pop\  with 
$\g(u)  = \bar \g$  and using that the angular momentum
$J= {\pi\ov 4 G_N} \bar \g$  should take integer values as discussed 
after eq.\pop. 

In  general, the r.h.s. of the relation \yiy\ will
contain contributions of all possible near-horizon  perturbations in all 
 transverse directions, but as clear from the above discussion
the  oscillations  in  the  four  `external' spatial   dimensions 
   have dominant statistical weight for  $w,m,\k \gg 1$.
Taking into account the superpartners of the four bosonic 
oscillation directions 
 the   leading term in the statistical entropy 
is thus given by (cf. \sta,\rere) 
\eqn\stuy{
S_{stat} = 2\pi \sqrt {N_{L}^{(\g)}} = 2\pi \sqrt{wm\k}
 ={ \pi^2 \ov 2G_N} \sqrt {Q_1Q_2P}  \ ,  }
i.e. reproduces  the BH entropy \entr. 

In above discussion we 
were assuming that the world sheet theory is 
$(1,1)$ supersymmetric as in   
type II theory. The same  conclusions apply  to
 the case of the  heterotic string 
since we may assume the `symmetric' embedding of   the 5-brane solution \chs\ 
(i.e. with the spatial magnetic part of the spin connection 
being identified with the gauge field  background). 
Then the relevant 4-dimensional part of the heterotic \sm becomes identical to
that of type II theory.\foot{The  embedding of the time-like `fundamental string' components 
of the spin connection
into the gauge group is not  possible, and, indeed,   is not 
needed  for exact conformal invariance \HT.}

All the steps  leading to statistical derivation
of the $D=5$ black hole  entropy  \stuy\  have 
straightforward  generalisation to the case of the $D=4$
dyonic black hole  considered in  \refs{\US,\USS, \TT}. The corresponding conformal \sm 
is very  similar to \lag, i.e. 
   has  again the non-trivial 
transverse part represented by the ($(4,4)$ supersymmetric) 
4-dimensional model  (cf. \lag)
\eqn\ytyu{ L_\bot = (g_{mn} +   B_{mn })(x)  \del x^m \bd x^n  
+   {\cal R}\p(x)  } $$
= f(x)k(x)  \big[ \del x_4 + a_s (x) \del x^s\big] \big[ 
\bd x_4  + a_s (x) \bd x^s\big]   
 +  f(x)  k^{-1} (x) \del x^s \bd x^s $$ $$ + \ 
  b_s (x) (\del x_4 \bd x^s - \bd x_4 \del x^s)
  +   {\cal R}   \p(x) \  ,  $$
$$  \del_{p} b_q - \del_q b_{p}
 =- \epsilon_{pqs} \del^s   f   , \ \ \ 
\del_{p} a_q - \del_q a_{p}
 = -\epsilon_{pqs} \del^s   k\inv , \ \  \ \  p,q,s=1,2,3 \ , 
$$ $$ f = 1+{P_1\over r}\ ,\ \ \  \ \ 
k\inv = 1+{P_2\over r}\ ,  \ \ \ \ \  \p= \ha \ln f \ . $$
Here $x_4$ is a periodic coordinate and  $r^2 =x_sx_s$. $F$ and $K$  in \lag\ 
are now given by  $F\inv = 1 + { Q_1 \ov r}, \ K = { Q_2 \ov r}$.
The short-distance limit of the 
 $D=6$ conformal model is again 
described    by a  WZW model with level $\k= 4 P_1P_2/\a'$.
Dimensional reduction along $x_4$ and the string direction $y_5$ 
leads to extremal dyonic black hole  with the 
 corresponding BH entropy   
\eqn\yerw{
S_{BH} =  {\pi \ov  G_N}\sqrt{ Q_1 Q_2 P_1 P_2} 
= {  2 \pi }  \sqrt {wm \k } \ , }
which is  reproduced by counting supersymmetric marginal 
perturbations in the 
{\it four} relevant `external' directions $x_s,x_4$.
 Contributions of oscillations in other compact internal directions 
are subleading,   
in complete analogy with the $D=5$ case.

In conclusion, it is clear 
that there are many  similarities between 
the  present NS-NS conformal \sm approach  and the approach based 
on R-R  embedding and D-brane count of BPS states
(especially in the formulation proposed in \susm\ 
where the effective central charge is fixed).
We believe  that establishing a  direct translation 
between the two approaches  will be 
  illuminating for better understanding 
solitons and black holes in string theory. 

\bigskip
 {\bf Acknowledgements}

\noindent 
I am grateful to M. Cveti\v c, I. Klebanov and F. Larsen
for stimulating  discussions and correspondence.
This work was supported by 
 PPARC,
ECC grant SC1$^*$-CT92-0789 and NATO grant CRG 940870.

\vfill\eject
\listrefs
\end